\documentclass[epj]{svjour}
\usepackage{graphics}
\begin{document}
\title{Total and Differential Cross Sections for the pp$\rightarrow$pp\textbf{$\eta^\prime$}
 Reaction Near Threshold}
\author{
A. Khoukaz\inst{1,}\thanks{\emph{email: khoukaz@uni-muenster.de}} \and
 I. Geck\inst{1} \and
 C. Quentmeier\inst{1} \and
 H.-H. Adam\inst{1} \and
 A. Budzanowski\inst{2} \and
 R. Czy\.zykiewicz\inst{3} \and
 D. Grzonka\inst{4} \and
 L. Jarczyk\inst{3} \and
 K. Kilian\inst{4} \and
 P. Kowina\inst{4}\inst{,5} \and
 N. Lang\inst{1} \and
 T. Lister\inst{1} \and
 P. Moskal\inst{3}\inst{,4} \and
 W. Oelert\inst{4} \and
 C. Piskor-Ignatowicz\inst{3} \and
 T. Ro\.zek\inst{5} \and
 R. Santo\inst{1} \and
 G. Schepers\inst{4} \and
 T. Sefzick\inst{4} \and
 S. Sewerin\inst{4} \and
 M. Siemaszko\inst{5} \and
 J. Smyrski\inst{3} \and
 A. Strza{\l}kowski\inst{3} \and
 A. T\"{a}schner\inst{1} \and
 P. Winter\inst{4} \and
 M. Wolke\inst{4} \and
 P. W\"{u}stner\inst{4} \and
 W. Zipper\inst{5}
}                     
%
%
\institute{
Institut f\"{u}r Kernphysik, Westf\"{a}lische Wilhelms-Universit\"{a}t, D-48149 M\"{u}nster, Germany
 \and
Institute of Nuclear Physics, Pl-31-342 Cracow, Poland
 \and
Institute of Physics, Jagellonian University, PL-30-059 Cracow, Poland
 \and
IKP and ZEL, Forschungszentrum J\"{u}lich, D-52425 J\"{u}lich, Germany
 \and
Institute of Physics, University of Silesia, PL-40-007 Katowice, Poland}
\date{Received: date / Revised version: date}
%
\abstract{
The $\eta^\prime$ meson production in the reaction $pp\rightarrow
pp\eta^\prime$ has been studied at excess energies of Q = 26.5,
32.5 and 46.6 MeV using the internal beam facility COSY-11 at the
cooler synchrotron COSY. The total cross sections as well as one
angular distribution for the highest Q-value are presented. The
excitation function of the near threshold data can be described by
a pure s-wave phase space distribution with the inclusion of the
proton-proton final state interaction and Coulomb effects. The
obtained angular distribution of the $\eta^\prime$ mesons
is also consistent with pure s-wave production.
} 
\maketitle
\section{Introduction}
\label{intro}
Measurements on the production of $\eta^\prime$ mesons, the
heaviest representative of the multiplet of pseudoscalar mesons,
allow to study the properties and the structure of this iso-scalar
meson, which are still far from being well known. Since states
with the same quantum numbers IJ$^P$
can mix, the physically observable
particles $\eta$ and $\eta^\prime$ are considered to be
mixed states of the I = 0 members of the ground state pseudoscalar
octet and singlet, commonly denoted as $\eta_8$ and $\eta_1$.
In case of an ideal mixing, the $\eta$ meson would have a pure
non-strange content ($u\bar{u}+d\bar{d}$), while the $\eta^\prime$
would show up as a pure $s\bar{s}$ state, corresponding to a
mixing angle of $\theta_{ideal} = - \arctan \sqrt{2}\, \approx
\,-54.7^\circ$. This value is in contrast to the experimentally
still inaccurately determined mixing angle $\theta_P$, which has been subject of
several investigations
\cite{Isg76,Kaz76,Dia81,Don85,Gil87,Sch93,Bal96,Jak96,Bra97,Bur98,Gua99,Mcn00}
and was found to be between $-9$ and $-20^{\circ}$.\\
Furthermore, studies on the production of $\eta^\prime$ mesons are
also important with respect to still controversially discussed
topics like possible $c\bar{c}$ or gluonic components in the
structure of the $\eta^\prime$ meson
\cite{Jak96,Hal97,Che97,Atw97,Nik98,Fel00,x1,x2} or the
understanding of the unexpected high mass, which is discussed in
the context of the U(1)$_A$ anomaly \cite{Fel00,Hoo76,x3}.
The gluonic contribution to the cross section for the
$NN\rightarrow NN\eta^\prime$ reaction may be inferred by the
comparison of the $\eta^\prime$ production in different isospin
channels \cite{x1,x2,x4}.\\
Recently, detailed measurements on the $\eta^\prime$ meson
production in the reaction channel $pp\rightarrow pp\eta^\prime$
have been performed in the previously unexplored region close to
threshold up to an excess energy of Q = 24 MeV
\cite{Hib98,Mos98,Mos00,Mos00b} as well at a higher excess energy
of Q = 144 MeV \cite{Bed98}. In this publication we present new
results on this reaction channel at intermediate excess energies
of Q = 26.5, 32.5 and 46.6 MeV, filling the gap
between the available data sets.\\
Due to the small relative momenta of the ejectiles in the region
of low excess energies, only partial waves of the lowest order
participate in the exit channel. Therefore, total and differential
cross section data yield nearly unscreened information on relevant
production mechanisms and allow to study final state interactions
(FSI) of the participating particles. Consequently, these new data
became subject of several model calculations and comparisons with
the related reaction channels on the $\pi^0$
and $\eta$ meson production \cite{x5,x6,x7}.\\
In contradistinction to the corresponding $\eta$ meson production
channel, whose production amplitude is assumed to be dominated by
the excitation of the $S_{11}$ nucleon resonance $N^*(1535)$, the
relevance of possible production diagrams to describe the
$\eta^\prime$ meson formation is still controversially discussed.
A prediction for both the shape and the absolute scale of the near
threshold excitation function of the $\eta^\prime$ production via
the reaction $pp\rightarrow ppX$ has been determined by Hibou et
al. \cite{Hib98}, comparing the $pp\eta$ and $pp\eta^\prime$
channels within an one-pion exchange model and adjusting an
overall normalization factor to fit the $pp\eta$ total cross
section data. The obtained prediction for the shape of the
$\eta^\prime$ excitation function is able to describe the data
well. However, the absolute scale of the total cross sections is
underestimated by a factor of 2-3 by these calculations, which
might be interpreted as a signal for the relevance of exchange
diagrams of heavier mesons (e.g. $\rho$) \cite{Hib98} or the
importance of
the gluonic contact term in the $\eta^\prime$ production \cite{x2}.\\
Contrary, model calculations by Sibirtsev et al. \cite{Sib98},
also based on the one-pion exchange diagram including the
proton-proton final state interaction, have been found to be able
to describe both the shape as well as the absolute scale of the
near threshold total cross section data. This result is argued to
indicate either only negligible contributions of the exchange of
heavier mesons or a mutual cancellation of their contributions. It
should be noted that in both models contributions of initial state
interactions of both protons have been neglected, which have been
reported to scale the absolute size of the total cross sections by
a factor of f = 0.2 \cite{x8} and f = 0.33 \cite{Nak99}
in the near threshold region for the $\eta$ and $\eta^\prime$ mesons,
respectively.\\
Recently, the $\eta^\prime$ meson production has been investigated
theoretically by Naka\-yama et al. \cite{Nak99} within a
relativistic meson exchange model, considering the exchange of
$\pi$, $\eta$, $\rho$, $\omega$, $\sigma$, $a_0$ mesons and
including effects of the proton-proton initial and final state
interaction. These calculations include contributions of nucleonic
and mesonic currents as well as contributions of nucleon
resonances, denoted as  S$_{11}$(1897) and P$_{11}$(1986), which
have been observed in multipole analyses of $\eta^\prime$
photoproduction experiments off protons \cite{Plo98}. Though the
nucleonic and the mesonic currents are found to reproduce the
observed cross sections, also contributions of the S$_{11}$(1897)
resonance alone are reported to be sufficient to describe the
data. However, to determine the relative magnitude of these
currents, total and differential cross section data at higher
excess energies are needed.

\section{Experiment}
\label{sec:1}
\begin{figure}
\resizebox{0.47\textwidth}{!}{%
\includegraphics{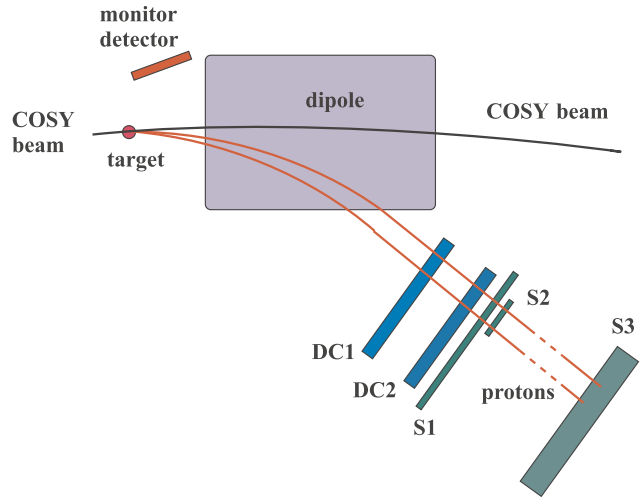}
}
\caption{Sketch of the internal beam facility COSY-11.}
\label{c11}       
\end{figure}

\begin{figure}
 \resizebox{0.47\textwidth}{!}{%
 \includegraphics{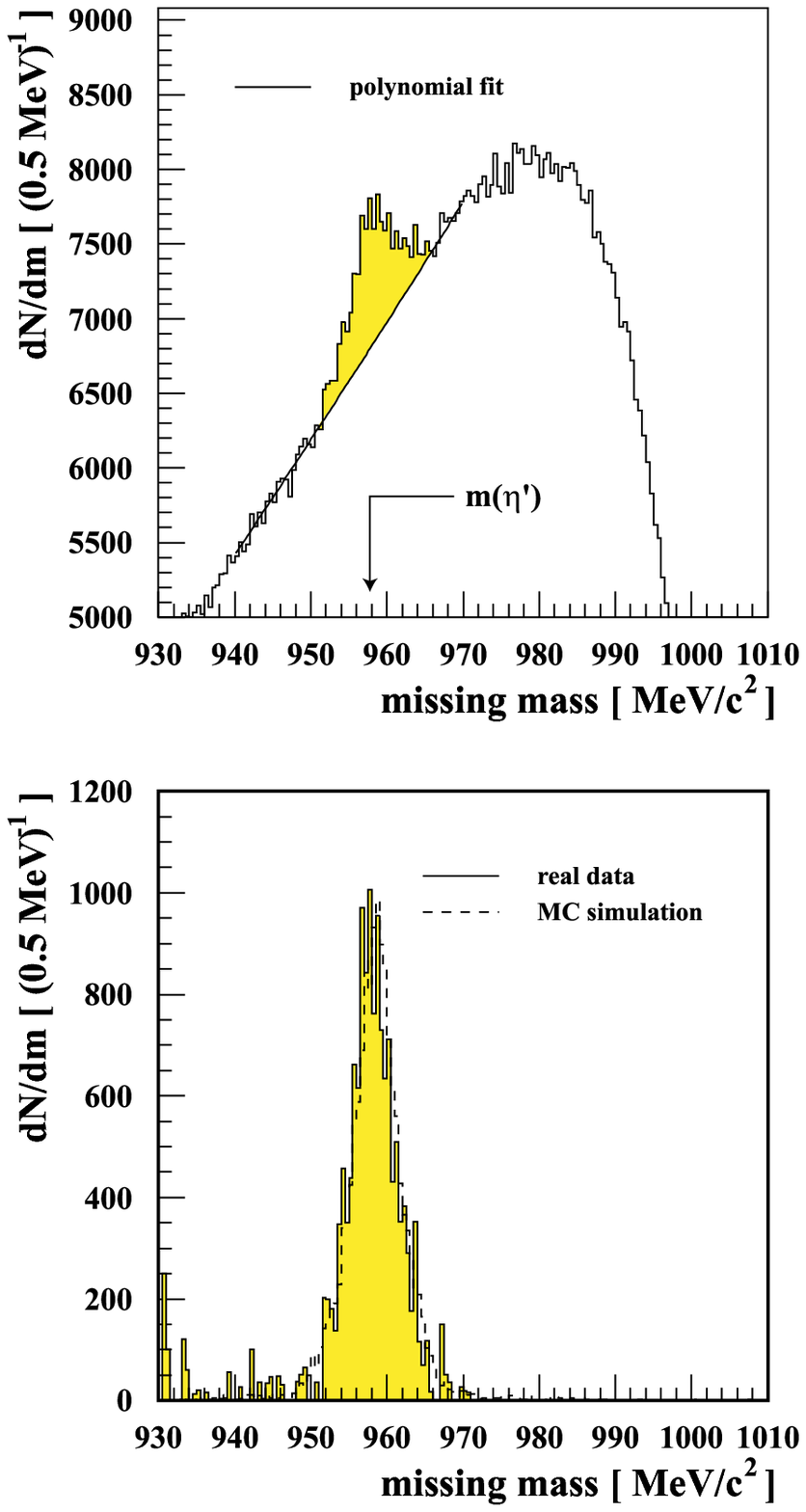}}
 \caption{Missing mass distribution of the selected two track events
 with both particles identified as protons (Q = 46.6 MeV). The lower figure
 presents the experimental data after subtraction of the background (see upper
 figure). In addition, the resulting $\eta^\prime$ signal (solid line) is compared
 with expectations according to Monte-Carlo simulations (dashed line).}
 \label{komb}
 \end{figure}

\begin{figure}[hb]
 \resizebox{0.49\textwidth}{!}{%
 \includegraphics{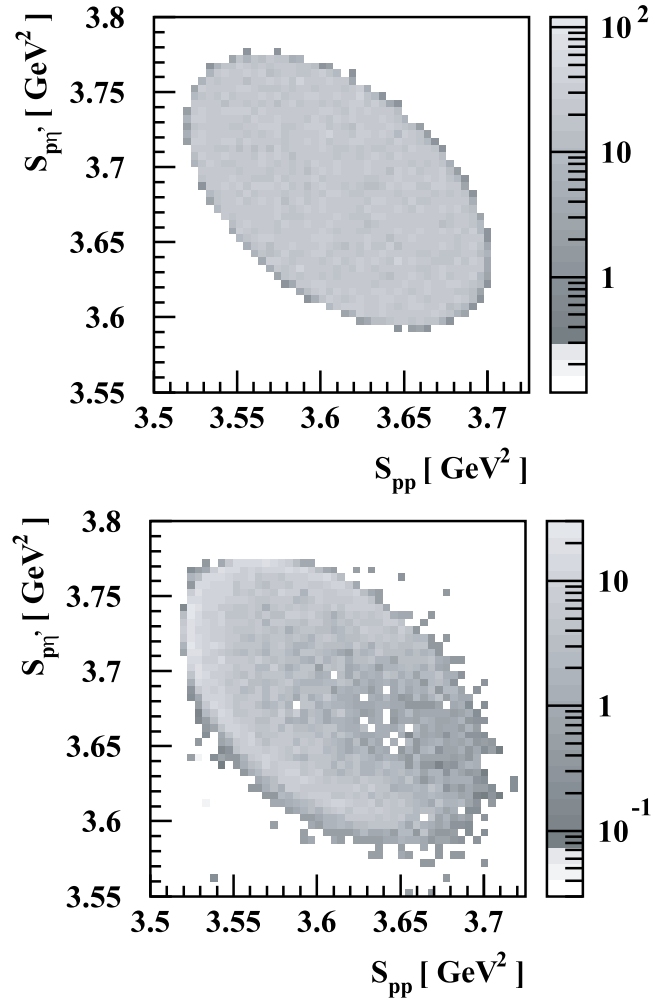}}
 \caption{Dalitz plots for generated (upper figure) and by the COSY-11
 detection system accepted (lower figure) Monte-Carlo events of the reaction
 $pp\rightarrow pp\eta^\prime$ at an excess energy of Q = 46.6 MeV.}
 \label{dalitz}
 \end{figure}

Measurements on the reaction $pp \rightarrow pp\eta^\prime$ have
been performed at the internal beam facility COSY-11 \cite{Bra96}
at COSY-J\"{u}lich \cite{Bec96}, using a hydrogen cluster target
\cite{Dom97} in front of a COSY-dipole magnet, acting as a
magnetic spectrometer. Tracks of positively charged particles,
detected in a set of two drift chambers (DC1 and DC2, fig.
\ref{c11}), can be traced back through the magnetic field to the
interaction point, leading to a momentum determination.
The
velocities of these particles are accessible via a time-of-flight
path behind the drift chambers, consisting of two scintillation
hodoscopes (start detectors S1 and S2) followed by a large
scintillation wall (S3) at a distance of $\sim$ 9.3 m, acting as a
stop detector. By measuring the momentum and the velocity,
particles are identified via invariant mass, i.e. the four
momentum vectors $P_i$
of positively charged ejectiles are fully determined.\\
The event selection for the reaction $pp \rightarrow
pp\eta^\prime$ was performed by accepting events with two
reconstructed tracks in the drift chambers, requiring both
particles being identified as protons. The four-momentum
determination of the positively charged ejectiles yields a full
event reconstruction for the reaction type $pp\rightarrow ppX$ and
allows an identification of the X-particle using the missing mass
method
 \begin{equation}
  m_x = |P_{beam}+P_{target}-P_{proton1}-P_{proton2}|
 \end{equation}
and to study angular distributions of the ejectiles. This
situation is demonstrated in fig. \ref{komb} for a beam momentum
of $p_{beam}$ = 3.356 GeV/c, corresponding to an excess energy of
Q = 46.6 MeV above the $\eta^\prime$ meson production threshold.
In the raw spectrum (upper figure) a signal of the $\eta^\prime$
meson production is clearly visible on a background arising from
multi pion production channels. The lower spectrum presents the
$\eta^\prime$ missing mass peak with a content of N $\sim$ 13000
events after subtraction of the background, which was fitted by a
first order polynomial. The observed mean position of the missing
mass peak, m$_X$\,=\,958.1 MeV/c$^2$, differs by 0.3 MeV/c$^2$ from
the nominal value (m$_{\eta^\prime}$ = 957.78 $\pm$ 0.14 MeV/c$^2$
\cite{Pdb}), reflecting the precision of the experimental method
and the quality of the accelerator beam. The missing mass
resolution amounts to $\sigma$ = 3 MeV/c$^2$
($\Gamma_{\eta^\prime}$ = 0.202 $\pm$ 0.016 MeV/c$^2$ \cite{Pdb}),
consistently with
Monte-Carlo simulations (fig. \ref{komb}, dashed line) based on GEANT 3.21 \cite{Gea93}.\\
To extract total and differential cross section data it is
important to investigate the phase space coverage of the detection
system. For the highest energy data point presented in this paper
this situation is illustrated in fig.~\ref{dalitz} presenting
the squared invariant mass of the proton-$\eta^\prime$ system
S$_{p\eta^\prime}$ as a function of the squared invariant mass of
the proton-proton system S$_{pp}$
 for Monte-Carlo events\footnote{The squared invariant mass S$_{ij}$
 of two particles $i$ and $j$ with
the four-momentum vectors $P_i$ and $P_j$ is given by S$_{ij}$ =
$|P_i+P_j|^2$.}.

As expected, the Dalitz plot displaying all
generated events (upper figure) is homogeneously filled, while
the corresponding plot for accepted and reconstructed events
(lower figure) presents an inhomogeneous structure, reflecting
the acceptance of the COSY-11 detection system.
However, from the
latter spectrum it is obvious that even at an excess energy of Q =
46.6 MeV the
whole Dalitz plot is covered.
The overall detection efficiencies, requiring the detection of
both protons, were determined to be\\
$\varepsilon$(Q = 26.5 MeV) = (2.0$^{+0.4}_{-0.3})\times$10$^{-2}$,\\
$\varepsilon$(Q = 32.5 MeV) = (1.5$^{+0.4}_{-0.3})\times$10$^{-2}$, and\\
$\varepsilon$(Q = 46.6 MeV) = (9.2$^{+3.2}_{-2.1})\times$10$^{-3}$.\\
Especially at high energies the uncertainty in the determination of the
detection efficiency is mainly given by application of different models for the proton-proton FSI
\cite{Dru97,Mor68,Nai77,Noy71,Noy72}.\\
To receive an $\eta^\prime$ meson angular distribution,
the range of scattering angles in the center of mass system
was divided into eight angular bins and missing mass spectra
have been extracted for each bin.
The contents of the missing mass peaks have been acceptance
corrected by results from phase space
Monte-Carlo simulations including the pp FSI, according to
\cite{Dru97,x9}. The inclusion of the final state interaction itself in the
event generator is motivated by the use of overall detection efficiencies
in the analysis.

\section{Results}
\label{sec:2}
%

\begin{figure}[h]
 \resizebox{0.49\textwidth}{!}{%
 \includegraphics{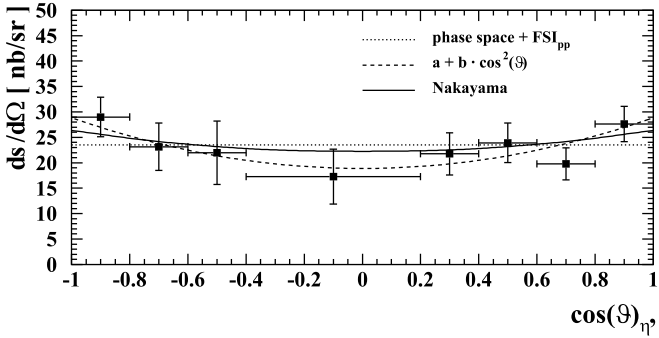}}
 \caption{Angular distribution of the emitted $\eta^\prime$ meson
 in the center of mass system at an excess energy of Q = 46.6 MeV.
 Newest results from \cite{Nak03} are represented by the solid line.
 For comparison, the dashed and the dotted lines indicate expectations
 according to phase space considerations with and without the inclusion
 of a $\cos^2(\vartheta)$ term.}
 \label{winkel}
 \end{figure}

\begin{figure*}
 \resizebox{.99\textwidth}{!}{%
 \includegraphics
 {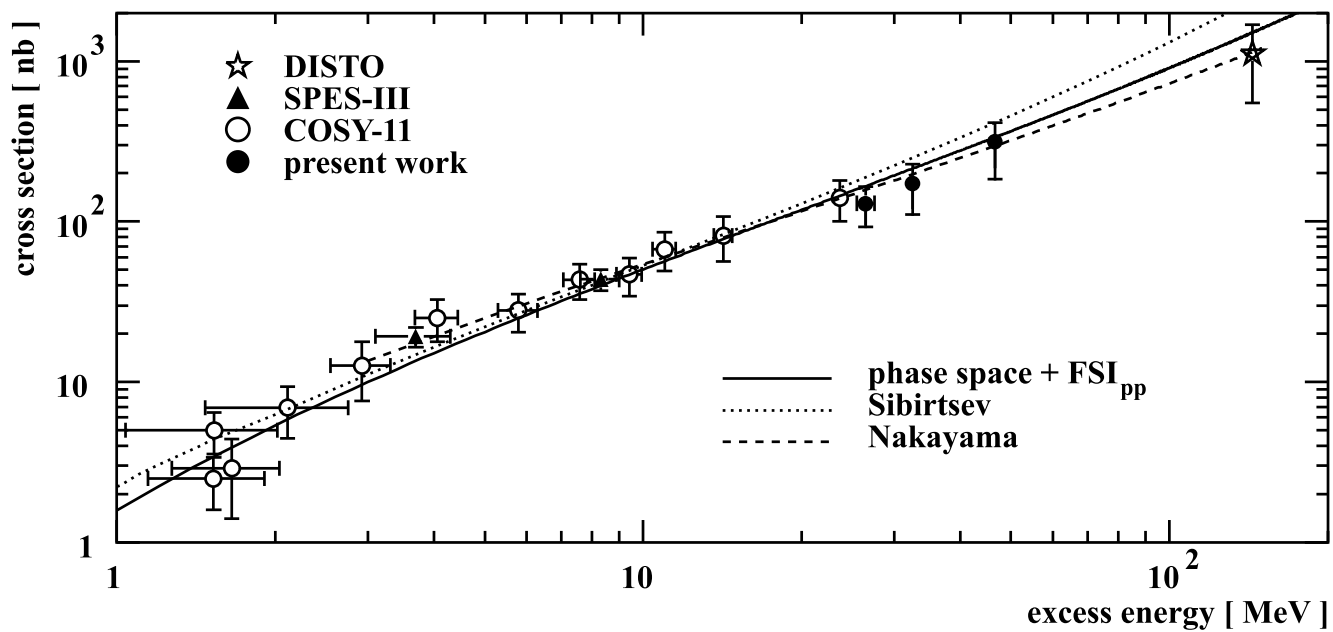}}
 \caption{Total cross sections for the reaction $pp\rightarrow pp\eta^\prime$
 as a function of the excess energy Q. Filled circles correspond to
 the data presented in this paper and triangles, open circles and
 stars correspond to data from \cite{Hib98,Mos98,Mos00,Bal00}, respectively.
 The curves are explained in the text.}
 \label{sigma}
 \end{figure*}

In fig. \ref{winkel} the resulting angular distribution of the
emitted $\eta^\prime$ mesons in the overall center of mass system
is presented for an excess energy of Q = 46.6 MeV.
The quoted
errors include statistical and systematical errors except
contributions from overall systematical uncertainties (e.g.
luminosity determination). The differential cross sections are
compatible ($\chi^2$ = 0.92) with an isotropic emission (dotted line),
indicating a
dominance of S-waves in the final state, consistent with results
obtained from the DISTO collaboration at an excess energy of Q =
144 MeV \cite{Bal00}. However, fig. \ref{winkel} might also
indicate contributions of higher partial waves, i.e. D-waves.
An inclusion of a $\cos^2(\vartheta)$ term to account for higher partial waves
leads to an adequate description of the angular distribution as demonstrated by the dashed line
($\chi^2$ = 0.45).\\
The integrated luminosities have been determined by comparing the differential
counting rates of elastically scattered protons with data
obtained by the EDDA collaboration \cite{Alb97}.
For beam momenta of p = 3.292 GeV/c, 3.311 GeV/c and 3.356 GeV/c,
integrated luminosities of  $\int L\,dt$ = 908 nb$^{-1}$, 841
nb$^{-1}$ and 4.50 pb$^{-1}$ (all: $\pm$ 1\% (stat.) $\pm$ 5\% (syst.))
have been determined.\\
The obtained values of total cross sections are listed in table
\ref{tab:1}. The overall systematical error arises from the
detection efficiency determination, the calculation of the
luminosity, the uncertainty of the COSY beam momentum ($\Delta$p/p
$\approx$ 0.1\%) as well as from acceptance corrections of the differential cross
sections.\\
\begin{table}
\caption{Total cross sections for the reaction $pp\rightarrow
pp\eta^\prime$. }
\label{tab:1}       
\begin{tabular}{rcl}
\hline\noalign{\smallskip}
excess energy  & & absolute cross section  \\
Q $[$MeV$]$    & &  $\sigma$ $[$nb$]$      \\
%
\noalign{\smallskip}\hline
  & &       \\
  26.5 $\pm$ 1.0   & & 130.0   $\pm$   13.8 (stat.)      $^{+21.2}_{-24.8}$  (syst.)     \\
    \\
  32.5 $\pm$ 1.0   & & 174.1   $\pm$   20.2 (stat.)      $^{+34.3}_{-45.8}$  (syst.)     \\
    \\
  46.6 $\pm$ 1.0   & & 314.9   $\pm$   17.3 (stat.)      $^{+81.9}_{-116.5}$  (syst.)     \\
 & &       \\
 \noalign{\smallskip}\hline
\end{tabular}
\end{table}
%
In fig.~\ref{sigma} the results (filled circles) are compared with
existing data. The solid line represents an s-wave phase space
calculation (meson production matrix element $|M_{0}|^2$ = const.)
modified by the proton-proton final state interaction (FSI) and
Coulomb effects, scaled to fit the data \cite{x5}:
\begin{equation}
  \sigma \propto \int_{0}^{q_{max}} k_{NN} q^2 |M_{0}|^2 \cdot |M_{FSI,Coulomb}|^2 dq.
\end{equation}
In this notation $q$ and $k_{NN}$ represent the momentum of the
meson in the CMS and the momentum of either nucleon in the rest
frame of the NN-subsystem. Within the experimental errors this fit
is able to describe the whole set of existent data. Therefore, one
can conclude that no further assumptions like a significant
$\eta^\prime$-proton final state interaction are needed in order
to describe the excitation function. In particular, distinct
effects of higher partial waves can be rejected due to the
isotropy of the presented angular distribution and the one from
DISTO at Q = 144 MeV
\cite{Bal00}.\\
%
The dotted curve of fig.~\ref{sigma} represents calculations from Sibirtsev et al.
\cite{Sib98} based on a one-pion exchange diagram including the pp
final state interaction. While for excess energies below Q = 25 MeV the
observed excitation function is described well, the new data from
COSY-11 (filled circles) as well as
the cross section from DISTO are somewhat overestimated.\\
Additionally, fig.~\ref{winkel} and fig.~\ref{sigma} show the
newest calculation from Nakayama \cite{Nak03} for the near
threshold $\eta^\prime$ meson production in proton-proton
collisions (dashed lines), based on a relativistic meson exchange
model. Since the relative strengths as well as the absolute scales
of the considered mesonic, nucleonic and nucleon resonance
currents are not known, different combinations reproducing the
available total cross sections are possible. However, in a
combined analysis of the $\eta^\prime$ meson production in pp and
$\gamma$p interactions Nakayama succeeded to describe both
reactions within his model consistently \cite{Nak03}. In this
approach the contribution of the mesonic exchange current has been
fixed by the photoproduction data and appeared to be much smaller
than assumed in earlier calculations \cite{Nak99}. On the
contrary, contributions of at least an S$_{11}$ nucleon resonance
in the mass region of 1650 MeV/c$^2$ and probably a P$_{11}$
resonance in the mass region of 1880 MeV/c$^2$ were found to be
necessary in order to describe the energy dependence of the total
cross sections of both the $\gamma$p an pp data. Furthermore, in
proton-proton collisions contributions of the nucleonic exchange
current were estimated to be comparatively small in order to
describe within the given uncertainties both the excitation
function a well as the observed angular distributions of emitted
$\eta^\prime$ mesons presented in \cite{Bal00} and in this work
(fig.~\ref{winkel}).

\section{Summary}
\label{sec:3}
At the COSY-11 facility the near threshold $\eta^\prime$ meson
production in the reaction channel $pp \rightarrow pp\eta^\prime$
has been studied at excess energies of Q = 26.5, 32.5 and 46.6
MeV. The obtained total cross section data fill the region of
intermediate excess energies between the low energy data
\cite{Hib98,Mos98,Mos00,Mos00b,Bed98} and one data point at a high
excess energy \cite{Bal00}. It was demonstrated that within the
quoted errors the complete available excitation function can be
adequately described by calculations on basis of the three body
phase space behaviour including effects of the pp FSI only, and no
distinct contributions from the $\eta^\prime$-proton interaction
or higher partial waves are necessary for the interpretation of
the data.\\
A comparison of the data with a pion exchange model calculation
from Sibirtsev et al. \cite{Sib98} results in an overestimation of
the data presented in this paper. This observation is in agreement
with the results from
DISTO \cite{Bal00} at an excess energy of Q = 144 MeV.\\
%
Newest calculations of Nakayama \cite{Nak03}, based on a
relativistic meson exchange model, succeed to describe both the
available total cross section data as well as the observed angular
distribution of emitted $\eta^\prime$ mesons. The necessity to
consider contributions from nucleon resonance currents in order to
describe the observed data might be interpreted as a signal for
the role of nucleon resonances for the $\eta^\prime$ production,
similarly to the $\eta$ meson case but on a lower scale.

\section{Acknowledgements}
\label{sec:4}
We thank K. Nakayama for valuable discussions on the mechanism of
the $\eta^\prime$ production. This research project was supported
in part by the BMBF (06MS881I), the Bilateral Cooperation between
Germany and Poland represented by the Internationales B\"{u}ro DLR
for the BMBF (PL-N-108-95) and by the Komitet Bada\'n Naukowych
KBN, the European Community - Access to Research Infrastructure
action of the Improving Human Potential Programm and by the FFE
grants (41266606 and 41266654) from the Forschungszentrum
J\"{u}lich.

%
%

\end{document}